\documentclass[aps,prl,reprint,groupedaddress]{revtex4-2}
\usepackage{graphicx}
\usepackage{amsmath}
\usepackage{amsfonts}
\usepackage{subfigure}
\begin{document}

% Use the \preprint command to place your local institutional report
% number in the upper righthand corner of the title page in preprint mode.
% Multiple \preprint commands are allowed.
% Use the 'preprintnumbers' class option to override journal defaults
% to display numbers if necessary
%\preprint{}

%Title of paper
\title{Arbitrary Time Thermodynamic Uncertainty Relation from Fluctuation Theorem}    
%Entropic Equality for the Precision of Current Fluctuation}

% repeat the \author .. \affiliation  etc. as needed
% \email, \thanks, \homepage, \altaffiliation all apply to the current
% author. Explanatory text should go in the []'s, actual e-mail
% address or url should go in the {}'s for \email and \homepage.
% Please use the appropriate macro foreach each type of information

% \affiliation command applies to all authors since the last
% \affiliation command. The \affiliation command should follow the
% other information
% \affiliation can be followed by \email, \homepage, \thanks as well.
\author{Takaaki Monnai}
%\email[]{Your e-mail address}
%\homepage[]{Your web page}
%\thanks{}
%\altaffiliation{}
\affiliation{Department of Materials and Life Science, Seikei University, Tokyo, 180-8633, Japan}

%Collaboration name if desired (requires use of superscriptaddress
%option in \documentclass). \noaffiliation is required (may also be
%used with the \author command).
%\collaboration can be followed by \email, \homepage, \thanks as well.
%\collaboration{}
%\noaffiliation

\date{\today}

\begin{abstract}
The thermodynamic uncertainty relation (TUR) provides a universal entropic bound for the precision of the fluctuation of the charge transfer for example for a class of continuous time stochastic processes. However, its extension to general nonequilibrium dynamics is still an unsolved problem. 
In this Letter, we show TUR for an arbitrary finite time in terms of exchange fluctuation theorem applied to ensemble of copies of the original system by assuming a physical regularity condition for the probability distribution. 
As a nontrivial practical consequence, we obtain universal scaling relations among the mean and variance of the charge transfer in short time regime.  
In this manner, we can deepen our understanding on a link between two important rigorous relations, i.e., the fluctuation theorem and the thermodynamic uncertainty relation. 
\end{abstract}

% insert suggested keywords - APS authors don't need to do this
%\keywords{}

%\maketitle must follow title, authors, abstract, and keywords
\maketitle

% body of paper here - Use proper section commands
% References should be done using the \cite, \ref, and \label commands
%\section{}
% Put \label in argument of \section for cross-referencing
%\section{\label{}}
%\subsection{}
%\subsubsection{}
 {\it Introduction.---} 
In recent years, the development of nanotechnology has made it possible to rather freely manipulate small systems such as the rectification of current and power generation in nanojunctions\cite{Hartmann1,Fujisawa2}.
Therefore, it is of fundamental importance to investigate the operating principles that small systems universally follow.
Then, a natural question that arises is how the notion of the thermodynamics that provides an operating principle for macroscopic systems can be extended to nonequilibrium small systems?
This is actually one of the major unsolved problems of the nonequilibrium statistical mechanics. 
In this context, the recently developing stochastic thermodynamics provides a comprehensive framework of the first and the second laws\cite{Sekimoto1,Seifert1,Seifert2} in terms of the intrinsic fluctuation of the work and heat, the current, and the entropy production for small system sizes. 

The universal theorems that rigorously hold even in nonequilibrium systems are especially valuable.  
In particular, the fluctuation theorems (FTs) provide a class of model independent symmetry of the probability distribution of the entropy production\cite{Evans1,Gallavotti1,Kurchan1,Lebowitz1,Crooks1,Jarzynski1,Gaspard1,Jarzynski2,Andrieux1,Andrieux2,Rao1,Esposito1}, which reproduce the second law, and the linear and nonlinear response relations\cite{Lebowitz1,Andrieux1,Andrieux2}. FTs have been verified in various nonequilibrium mesoscopic systems for example for a dragged colloidal particle in water\cite{Wang1}, the electron transports in quantum dots\cite{Utsumi1,Nakamura1}, the heat conduction in nanojunctions\cite{Saito1}, to name but a few. More recently, the thermodynamic uncertainty relation (TUR) attracts considerable attention as another class of universal theorem\cite{Seifert3,Gingurich1} that sets fundamental bounds to the precision of a fluctuating charge in terms of the entropy production. 
TUR claims that the square mean to variance ratio expressing the thermodynamic precision of a fluctuation of current-like quantity $J$ is upper bounded by half of the mean entropy production $\sigma$
\begin{align}
&\frac{\langle J\rangle^2}{{\rm Var}[J]}\leq\frac{\sigma}{2}, \label{TUR1}
\end{align} 
which was first derived for the continuous time stochastic processes\cite{Gingurich1,Gingurich2,Polettini1,Pietzonka1,Pietzonka2}, subsequently generalized to finite time\cite{Gingurich3,Pietzonka3} as reviewed in \cite{Horowitz1}, and also to the quantum systems\cite{Goold1,Hasegawa2,Friedmann1,Monnai1}.  
Thus, TUR admits a simple interpretation, i.e., a large entropy production inevitably occurs to suppress the fluctuation.     

The mutual relation between FT and TUR is non-trivial. 
Indeed, FT is an equality containing symmetry relations among all the cumulants, and inevitably concerns with the rare events that causes a negative entropy production. On the other hand, TUR is an inequality expressed by the first and second order cumulants of a charge transfer, and therefore focuses on typical events characterized by the mean and variance. 
Nevertheless, there are a few remarkable progresses to derive TUR from FT by neglecting a small term\cite{Pietzonka3} or modifying entropic bounds\cite{Timpanaro1,Hasegawa1}.  
%Let us briefly describe these achievements.      
In Ref. \cite{Pietzonka3}, TUR is shown for the short time limit by approximating the variance ${\rm Var}[J]$ with the mean square $\langle J^2\rangle$ from the evaluation of the large deviation function with various numerical verifications. 
In Ref. \cite{Hasegawa1}, TUR with an exponential entropic bound is derived, and (\ref{TUR1}) is reproduced in the linear response regime. 
Ref. \cite{Timpanaro1} provides a derivation of TUR with an entropic bound saturated by the minimal distribution which is, however, singular and is given as a combination of delta functions on two points. Then, the consequent entropic bound is slightly looser than the standard bound $\frac{\sigma}{2}$. 

In this Letter, we provide another contribution to this significant unsolved problem to connect rare events to typical ones by directly showing TUR (\ref{TUR1}) for the charge transfer $J$ during an arbitrary time $\tau$ on the basis of a geometric argument from the exchange fluctuation theorem (EFT)\cite{Jarzynski2} under a regularity condition for the fluctuation of the charge transfer. We require that the probability distribution of the charge transfer satisfies the large deviation principle\cite{Ellis1}, and the rate function locally obeys that of the central limit theorem in the vicinity of the mean value. 
We also show that the condition of the equality sign in (\ref{TUR1}) quite generally holds in a relevant short time limit, and as a practical consequence we derive a universal relation among the scaling exponents of the mean and variance of the charge transfer.  

{\it Set up.---}
In what follows, we describe our set up. 
EFT holds both for autonomous and externally driven systems under a time symmetric protocol. 
Therefore, the charge transfer collectively refers to the heat current flowing  between two objects and also to the work done under a time symmetric protocol in general.
For simplicity of notation, we consider the case of heat transfer. We can similarly explore the case of other charge transfers such as the work done.  
 
Let us recall the systems where EFT for heat exchange holds: We consider two large objects $A$ and $B$ that are initially disconnected and prepared in equilibrium states at different temperatures $T_A(=\frac{1}{k_B\beta_A})$ and $T_B(=\frac{1}{k_B\beta_B})$.  Here, $\beta_A$ and $\beta_B$ denote the inverse temperatures with $k_B$ being the Boltzmann constant. 
Then, at $t=0$ the two large objects start to interact until time $t=\tau$, and separate again. Let $J$ denote the energy  transfered from $A$ to $B$ during $\tau$. 
This energy transfer or heat current behaves stochastically depending on macroscopically uncontrolable precise of the initial state, and let $p_\tau(J)$ denote the probability distribution of $J$, which is non-Gaussian in general. We will use an abbreviated notation for the affinity $\Delta\beta=\beta_A-\beta_B$. Without loss of generality, hereafter $\Delta\beta$ is supposed to be positive.  

For the present set up, EFT for heat or energy transfer holds\cite{Jarzynski2} under the weak coupling condition
\begin{eqnarray}
&&\frac{p_\tau(J)}{p_\tau(-J)}=e^{\Delta\beta J}. \label{EFT1}
\end{eqnarray}
\\
{\it Fluctuation Theorem for Copies.---}
Let us introduce a notion of independent and identical copies to extract underlying features of the fluctuation of the charge transfer.       
We consider $N$ identical copies of the original system (for heat conduction, two large objects $A$, $B$, and if it exists a link between them), which are mutually noninteracting.     
Hereafter, the original system of interest is identified with the first copy.   

To show (\ref{TUR1}), we extend EFT (\ref{EFT1}) to the net charge transfer $J_{tot}=\sum_{k=1}^NJ_k$, where $J_k$ denote that for the k-th copy.  This generalization is straightforward from the additivity of the charge transfer. 

By increasing the number of copies $N\rightarrow\infty$, the probability distribution of the net charge transfer $J_{tot}$ follows the large deviation principle\cite{Ellis1} 
\begin{align}
&{}^\exists {\rm I}(J)=\lim_{N\rightarrow\infty}\frac{1}{N}\log p_\tau(J_{tot}=NJ), \label{largedeviation1}
\end{align}
where ${\rm I}(J)$ is the rate function, which is nonnegative and convex. 
Then, EFT for the ensemble of copies is expressed as a symmetry 
\begin{align}
&{\rm I}(J)-{\rm I}(-J)=-\Delta\beta J. \label{EFT2}
\end{align}    
\\
{\it Geometric Derivation of TUR.---}  
The following derivation of TUR (\ref{TUR1}) is based on a geometric argument in terms of EFT (\ref{EFT2}) and the  regularity condition for the probability distribution $p_\tau(J_1)$ meaning that the rate function is well-approximated by that of the central limit theorem near the mean value. The regularity condition holds for example if the convergence to the rate function in (\ref{largedeviation1}) is rapid \cite{Supplement1}. 
As the main result of this Letter, we will derive TUR without any modification to the entropic bound by restricting to the physically natural systems that satisfy such a regularity condition. The entropy production $\sigma=\Delta\beta\langle J\rangle$ is identified as the product of the affinity and the mean value of the current. 

Here, we sketch the outline of our derivation. 
\begin{figure}
\includegraphics[scale=0.5]{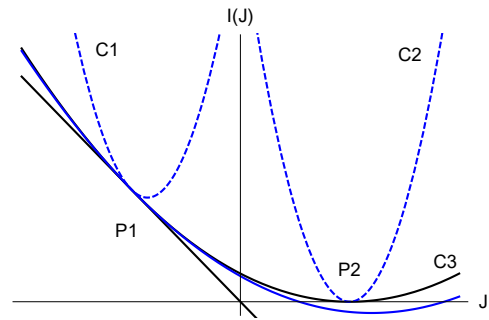}
\caption{Illustration of the curves ${\rm C}_k$ ($1\leq k\leq 3$). The curves ${\rm C}_1$ and ${\rm C}_2$ (dashed blue curve) have the same curvature from EFT which is supposed to be larger than that of ${\rm C}_3$. 
From the large deviation principle, ${\rm C}_1$ and ${\rm C}_2$ are tangent to the line $-\Delta\beta J$ and the horizontal axis, respectively. It turns out that ${\rm C}_1$ actually crosses the horizontal axis (blue curve).}
\end{figure}
 
Near the mean value $J=\langle J_1\rangle$, the rate function $I(J)$ is locally evaluated as a parabola  
\begin{align}
&{\rm I}(J)=\frac{(J-\langle J_1\rangle)^2}{2{\rm Var}[J_1]} \label{rate1}
\end{align}
from the central limit theorem. Let ${\rm C}_2$ denote the curve corresponding to (\ref{rate1}).  
    
We can also evaluate the rate function in the neighborhood of $J=-\langle J_1\rangle$ from EFT (\ref{EFT2}) and (\ref{rate1}) as another parabola ${\rm C}_1$ with the same curvature as in (\ref{rate1})
\begin{align}
&{\rm I}(J)=\frac{(J+\langle J_1\rangle)^2}{2{\rm Var}[J_1]}-\Delta\beta J. \label{rate2}
\end{align}
In this manner, the curve ${\rm C}_1$ and ${\rm C}_2$ provide restrictions to the rate function from rare and typical events, respectively.  

The outline of the derivation is that if TUR (\ref{TUR1}) does not hold, then  
a curve ${\rm C}$ corresponding to the rate function violates at least one of the conditions from the convexity, the central limit theorem, and EFT.   
We remark that TUR (\ref{TUR1}) is equivalent to a geometric condition that the curve ${\rm C}_1$ crosses the horizontal axis as a blue solid curve in Fig. 1. Actually, the nonnegativity of the discrimination of ${\rm C}_1$ is nothing but (\ref{TUR1}). 
Suppose TUR does not hold and the discrimination of ${\rm C}_1$ is negative. 
The conditions (\ref{rate1}), (\ref{rate2}), and the regularity condition imply that the curve ${\rm C}$ corresponding to the actual rate function is tangent to the line $-\Delta\beta J$ and the horizontal axis, and locally follows the curves ${\rm C}_1$ and ${\rm C}_2$ with a common curvature $\frac{1}{{\rm Var}[J_1]}$ at the points ${\rm P}_1=(-\langle J_1\rangle,\Delta\beta\langle J_1\rangle)$ and ${\rm P}_2=(\langle J_1\rangle,0)$. 
Such a curve ${\rm C}$, however,  is not convex and contradicts to the convexity of the rate function\cite{Supplement1}. This completes the derivation.      
 %Similarly, ${\rm C}_2$ is tangent to the point ${\rm P}_2=(-\langle J_1\rangle,\Delta\beta\langle J_1\rangle)$ with . 
\\
\\
Here, we compare our result with related works. 
If the equality condition is fulfilled in TUR (\ref{TUR1}) then the curves ${\rm C}_1$ and ${\rm C}_2$ coincide and form a common global curve ${\rm C}_3$, which is essentially equivalent to the quadratic bound used for a class of stochastic processes\cite{Gingurich1,Gingurich2,Gingurich3}. 
If the discrimination is zero, ${\rm C}$ corresponding to the rate function is given by ${\rm C}_3$.   
 
Note that our geometric argument can also make clear the mechanism of the violation of TUR for a class of irregular distributions $p_r(J)$ generated from the minimal distribution $p_{\rm min}(J)$ that has the smallest variance for a fixed mean value and is given as a sum of delta functions as reported in Ref. \cite{Timpanaro1}. 
In our case, the minimal distribution reads
\begin{align}
&p_{\rm min}(J)=\frac{e^{\frac{1}{2}\Delta\beta J}\delta(J-a)+e^{\frac{1}{2}\Delta\beta J}\delta(J+a)}{e^{\frac{1}{2}\Delta\beta a}+e^{-\frac{1}{2}\Delta\beta a}} \label{minimal1}
\end{align}
with an arbitrary positive constant $a$. 

The corresponding curve ${\rm C}_2$ has the largest curvature proportional to the inverse of the variance $\frac{1}{{\rm Var}[J]}$. 
Therefore, the curve ${\rm C}$ follows ${\rm C}_2$ only in a close vicinity of the mean value. As a consequence, the bound of TUR (\ref{TUR1}) is available only approximately with sufficiently small values of the affinity $\Delta\beta$.   
Furthermore, we can construct a class of TUR violating probability distributions $p_r(J)$ by generalizing (\ref{minimal1}) to continuous but finite supports. Such distributions are expressed as
\begin{align}
&p_r(J)=S_r(J)e^{\frac{1}{2}\Delta\beta J}, \label{minimal2}
\end{align}
where $S_r(J)(=S_r(-J))$ is an even function as in Ref. \cite{Timpanaro1} and importantly does not rapidly decay. 
For instance, it is easy to numerically verify that the choices $S_r(J)=1-(\frac{|J|}{a})^r$ with $r\geq 0$ also provide distributions which slightly violates the bound (\ref{TUR1}) for large $\Delta\beta$\cite{Supplement1} for the same reason as that of $p_{\rm min}(J)$, i.e., the smallness of variance due to the non-sufficient decay of $S_r(J)$. 
In contrast to the minimal distribution, the distributions for these choices of $S_r(J)$ are continuous. 
However, the support is exactly finite, which restricts the range of the fluctuation and is in marked contrast to the usual distributions of charge transfers with exponential tails.
Hence, (\ref{minimal2}) with the abovementioned choice of $S_r(J)$ is regarded as the onset of the regularity condition. \\   

{\it Short time regime.---}
%TUR (\ref{TUR1}) universally holds as an exact inquality for an arbitrary time $\tau$. 
In the remaining of the Letter, to investigate further universal model-independent properties, we will consider the short time limit.    

TUR applied to the short time limit claims that for the statistics of current, the mean $\langle J\rangle$ and the variance ${\rm Var}[J]$ satisfy the following theorem.    \\
\\
\textbf{Theorem}(Equality in a short time limit) \\
In a proper short time limit $\tau\rightarrow 0$, the ratio of the thermodynamic precision and the half of the entropy production approaches unity
\begin{eqnarray}
&&\lim_{\tau\rightarrow 0}\left(\frac{\langle J\rangle^2}{{\rm Var}[J]}\right)/\left(\frac{\sigma}{2}\right)=1.  \label{TPE1}
\end{eqnarray}
\\
The use of the short time limit is certainly a limiting and strong condition\cite{Pietzonka1,Timpanaro1}, however, we can apply (\ref{TPE1}) to the analysis of the universal relation among the scaling exponents for the charge transfer.  
As a nontrivial consequence of (\ref{TPE1}), we will show the following corollary later in (\ref{exponent1}). \\
\\
\textbf{Corollary}(Scaling of mean and variance in short time limit) \\
Let us independently vary the duration $\tau$ and the affinity $\Delta\beta$\cite{footnote1}.  
If the mean of the charge transfer follows the scaling relation in the short time regime, i.e., ${}^\exists p>0$ and ${}^\exists \alpha\geq 0$, 
\begin{eqnarray}
&&\langle J\rangle\propto \tau^p, \propto\Delta\beta^\alpha  \label{scaling1}
\end{eqnarray}
as $\tau\rightarrow 0$ then the variance scales as
\begin{eqnarray}
&&{\rm Var}[J]\propto \tau^p, \propto\Delta\beta^{\alpha-1} \label{scaling2}
\end{eqnarray}
and vice versa.

Note that in the short time regime, the affinity $\Delta\beta$ is not necessarily small. 
%Note that it is practical to regard $\tau$ and $\Delta\beta$ as independent variables e.g. by fixing the affinity $\Delta\beta$ and only changing $\tau$.   
As for the $\tau$ dependence, the validity of this corollary was experimentally verified for the electron current in a quantum dot\cite{Gustavsson1} in the context of the full counting statistics. 

{\it Derivation of the relation among the scaling exponents.---}
The central limit theorem states that for sufficiently large $N$ the probability distribution $p_\tau(J_{tot})$ obeys a normal distribution as
\begin{align}
&p_\tau(J_{tot})/\left({\cal C}e^{-\frac{(J_{tot}-N\langle J_1\rangle)^2}{2N{\rm Var}[J_1]}}\right)\cong 1 \label{distribution1}
\end{align}
for  
\begin{align}
&|J_{tot}-N\langle J_1\rangle|\leq \kappa_1\sqrt{N{\rm Var}[J_1]} \label{deviation1}
\end{align}
with a normalization constant ${\cal C}$ and an ${\cal O}(1)$ dimensionless parameter $\kappa_1$\cite{footnote2}. 
Fix $N$ large but finite, the range of applicability (\ref{deviation1}) contains $[-\langle J_1\rangle,\langle J_1\rangle]$  as the central region with a nonnegligible probability $p_\tau(J_{tot})$ by taking $\tau$ sufficiently short so that $N\langle J_{1}\rangle$ is kept finite.  

Then, a particular choice $|J_{tot}|=\kappa_2\langle J_{tot}\rangle$
 with $\kappa_2={\cal O}(1)$ fulfills the condition (\ref{deviation1}) in a proper short time regime. 
For this $J_{tot}$, we obtain
\begin{eqnarray}
&&\frac{p_\tau(J_{tot})}{p_\tau(-J_{tot})}\cong e^{2\frac{\langle J_1\rangle}{{\rm Var}[J_1]}J_{tot}} \label{ratio3}
\end{eqnarray} 
from (\ref{distribution1}). 
The main idea is to compare (\ref{ratio3}) with EFT applied to the probability distribution of the net charge transfer  
\begin{eqnarray}
&&\frac{p_\tau(J_{tot})}{p_\tau(-J_{tot})}=e^{\Delta\beta J_{tot}}. \label{EFT3}
\end{eqnarray} 

To explain the validity of this scenario, let us investigate the asymptotic scaling behaviors of the mean and variance by assuming the following ansatz in the short time limit
\begin{eqnarray}
&&\langle J_1\rangle=K_1(\frac{\tau}{\tau_0})^p(\frac{\Delta\beta}{\Delta\beta_0})^\alpha \label{scaling3}  
\end{eqnarray}
and
\begin{eqnarray}
&&{\rm Var}[J_1]=K_2(\frac{\tau}{\tau_0})^q(\frac{\Delta\beta}{\Delta\beta_0})^\gamma, \label{scaling4}
\end{eqnarray}
where the coefficients $K_1$, $K_2$ and standard values of the time scale $\tau_0$ and the affinity $\Delta\beta_0$ are constant. At the moment, the relation among the exponents $p$, $q$, $\alpha$, and $\gamma$ are unknown.   

The condition for the central limit theorem (\ref{deviation1}) requires that  
$|N\langle J_1\rangle|\leq \kappa_1\sqrt{N{\rm Var}[J_1]}$
holds. 
By substituting (\ref{scaling3}) and (\ref{scaling4}) into this condition, a straightforward calculation shows that 
the choice 
\begin{eqnarray}
&&p=q, \alpha=\gamma+1 \label{exponent1}
\end{eqnarray}
is a unique solution that satisfies (\ref{ratio3}), EFT (\ref{EFT3}), and (\ref{deviation1}).  
Interestingly, in the linear response regime, i.e., $\alpha=1$, $\gamma$ becomes vanishingly small.  
We verified (\ref{exponent1}) in concrete examples in Supplemental Material.

Combining EFT (\ref{EFT3}) and (\ref{ratio3}), we can show that the fluctuation of the heat current satisfies 
\begin{eqnarray}
&&\lim_{\tau\rightarrow 0}\frac{2\langle J_1\rangle}{{\rm Var}[J_1]}=\Delta\beta. \label{TPE3}
\end{eqnarray}
By multiplying the mean value $\langle J_1\rangle$, and using $\sigma=\Delta\beta \langle J_1\rangle$, equality of TUR (\ref{TPE1}) holds in the short time limit.

Similarly, we can obtain TUR  equality in the short time limit for the external work. 

Next, we will show a concrete example of TUR (\ref{TUR1}) and the equality (\ref{TPE1}) for Hamiltonian dynamics, which supports our perspective. 
\\
\\
{\it Example.---}
We consider a one-dimensional oscillator with a time-dependent frequency $\omega(t)$ as one of the simplest models with non-Gaussian probability distributions under the time symmetric driving protocol. 
We explore the work done during $\tau$ as a charge transfer.  
Let $q$ and $p$ denote the position and momentum of a particle with a mass $m$, and let 
\begin{eqnarray}
&&H(q,p,t)=\frac{p^2}{2m}+\frac{m\omega(t)^2}{2}q^2 \label{oscillator1}
\end{eqnarray}
denote the Hamiltonian at time $t$. 
For concreteness, the time dependence of the frequency is assumed to be time symmetric $\omega(\tau-t)=\omega(t)$ under the protocol  $m\omega(t)^2=m\omega_0^2(1+\frac{2t}{\tau})$ for $0\leq t\leq\frac{\tau}{2}$ and $m\omega(t)^2=m\omega_0^2(3-\frac{2t}{\tau})$ for $\frac{\tau}{2}\leq t\leq\tau$ with a natural frequency 
$\omega_0$.
Initially, the state $(q(0),p(0))$ is sampled from the canonical ensemble at a room temperature $T=300 {\rm K}$. 

Then, we can explicitly solve the equation of motion, and calculate the work $W=\int_0^\tau m\omega(t)\dot{\omega}(t)q(t)^2dt$. 
We illustrate the TUR ratio $\left(\frac{\langle W\rangle^2}{{\rm Var}[W]}\right)/\left(\frac{\sigma}{2}\right)$ as a function of $\tau$ in Fig. 2, and confirm TUR and (\ref{TPE1}). 
Since we are interested in mesoscopic systems, we fix the mass to $m=10^{-9} {\rm Kg}$ and change the spring constant $m\omega_0^2$ so that the period of oscillation is in the range between $0.07 {\rm s}$ to $0.3 {\rm s}$. 
As shown in Fig. 3, the work distribution has a sharp peak, however, the TUR ratio actually converges to unity in the short time regime, which is in agreement with (\ref{TPE3}).    
We can also verify that (\ref{scaling1}) and (\ref{scaling2}) holds with the common exponent $p=2$ and $\alpha=0$ by replacing $J$ and $\Delta\beta$ with $W$ and $\beta$, respectively\cite{Supplement1}.   

\begin{figure}
\includegraphics[scale=0.48]{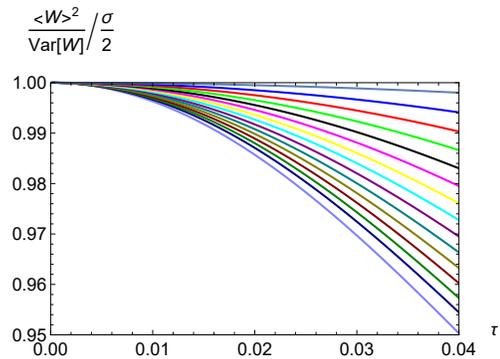}
\caption{Plot of the TUR ratio for the work done $\left(\frac{\langle W\rangle^2}{{\rm Var}[W]}\right)/\left(\frac{\sigma}{2}\right)$ (array of dots) against the time duration $\tau$. The horizontal axis is measured in units of seconds. We changed the value of the spring constant from $m\omega_0^2=10^{-8} {\rm Kg/s^2}$ (the top  light blue) to $m\omega_0^2=2.9\times 10^{-7} {\rm Kg/s^2}$ (the down light blue). 
}
\end{figure}
\begin{figure}
\includegraphics[scale=0.47]{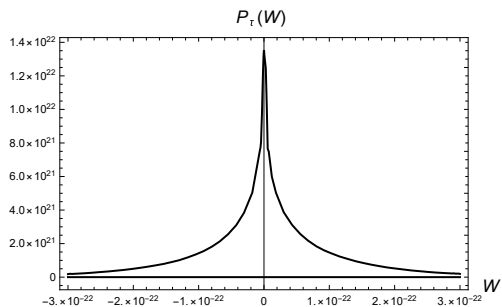}
\caption{The probability distribution of work done $p_\tau(W)$. The horizontal axis is measured in units of ${\rm J}$. We set the values of the spring constant $m\omega_0^2=10^{-8} {\rm Kg/s^2}$ and $\tau=0.02 {\rm s}$. The plot shows the non-Gaussian nature of the probability distribution.}
\end{figure}

{\it Conclusion.---} 
We derived TUR for the charge transfer by a geometric argument on the rate function in terms of EFT and the central limit theorem applied to the ensemble of independent copies of the original system.   
In particular, TUR has a simple geometric interpretation in terms of the discrimination of the curve ${\rm C}_1$, which locally characterizes the statistics of rare events. 
In this manner, we directly revealed a link between important rigorous theorems TUR and EFT under a physical requirement of the regularity condition,  which complements the insights gained in Refs. \cite{Pietzonka1,Hasegawa1,Timpanaro1}. Exclusion of irregular distributions generated by the minimal distribution has a drawback that restricts the availability of our result, however, has also an advantage that no modification is necessary to the standard entropic bound. Hence, our result guarantees that TUR (\ref{TUR1}) holds for a large class of practical dynamics starting from a local Gibbs ensemble\cite{Jarzynski2}.     

In the short time regime, TUR equality (\ref{TPE1}) holds beyond the linear response regime by fixing the affinity large but finite. If the net charge transfer quickly converges to the normal distribution\cite{footnote3,Callen1}, the TUR equality (\ref{TPE1}) would accurately hold without taking the limit $\tau\rightarrow 0$.

As a nontrivial corollary, we have shown a universal relation among the scaling exponents of the mean and variance of the charge transfer in the short time regime. 
This prediction is practically important e.g. for the electron transports in nanojunctions\cite{Gustavsson1}, for our example and for a stochastic description of a dragged colloidal particle in water\cite{Wang1}.  
The corollary provides a unified theoretical explanation for these observations. 
\begin{acknowledgments}
This work was supported by the Grant-in-Aid for Scientific Research (C) (No.~18K03467 and No.~22K03456) from the Japan Society for the Promotion of Science (JSPS).
% put your acknowledgments here.
\end{acknowledgments}

% Create the reference section using BibTeX:

%\bibliography{basename of .bib file}

\begin{thebibliography}{30}
\bibitem{footnote1}Note that it is practical to regard $\tau$ and $\Delta\beta$ as independent variables by fixing the affinity $\Delta\beta$ and changing $\tau$.
\bibitem{footnote2} Eq. (\ref{distribution1}) is a stronger statement than the stochastic convergence of the probability distribution $\lim_{N\rightarrow\infty}p_\tau(\frac{J_{tot}-N\langle J_1\rangle}{\sqrt{N{\rm Var}[J_1]}}=J)=\frac{1}{\sqrt{2\pi}}e^{-\frac{J^2}{2}}$, and is valid near the mean value $N\langle J_1\rangle$.
\bibitem{footnote3} In many cases, the convergence to the normal distribution in the central limit theorem is sufficiently rapid. The distribution function in Fig. 3 is an example.
\end{thebibliography}
\end{document}

% Use the \preprint command to place your local institutional report
% number in the upper righthand corner of the title page in preprint mode.
% Multiple \preprint commands are allowed.
% Use the 'preprintnumbers' class option to override journal defaults
% to display numbers if necessary
%\preprint{}

%Title of paper
\title{Supplemental Material}    
%Entropic Equality for the Precision of Current Fluctuation}

% repeat the \author .. \affiliation  etc. as needed
% \email, \thanks, \homepage, \altaffiliation all apply to the current
% author. Explanatory text should go in the []'s, actual e-mail
% address or url should go in the {}'s for \email and \homepage.
% Please use the appropriate macro foreach each type of information

% \affiliation command applies to all authors since the last
% \affiliation command. The \affiliation command should follow the
% other information
% \affiliation can be followed by \email, \homepage, \thanks as well.
\author{Takaaki Monnai}
%\email[]{Your e-mail address}
%\homepage[]{Your web page}
%\thanks{}
%\altaffiliation{}
\affiliation{Department of Materials and Life Science, Seikei University, Tokyo, 180-8633, Japan}

%Collaboration name if desired (requires use of superscriptaddress
%option in \documentclass). \noaffiliation is required (may also be
%used with the \author command).
%\collaboration can be followed by \email, \homepage, \thanks as well.
%\collaboration{}
%\noaffiliation

%\date{\today}
% insert suggested keywords - APS authors don't need to do this
%\keywords{}

%\maketitle must follow title, authors, abstract, and keywords
\maketitle
\begin{widetext}
\section*{Supplemental Material}
This Supplemental Material consists of four sections. In Sec. I, we complement some details of the proof of TUR 
In Sec. II, we provide details on the derivation of the equality for TUR in the short time limit, and explain the relations among the scaling exponents of the mean and variance of the charge transfer.
In Sec. III, we give additional numerical results on the example in the main text such as the scaling of the charge transfer in the short time regime and a verification of EFT. As another example, we describe an experimental design for an externally dragged Brownian particle. 
In Sec. IV, we give a derivation of EFT\cite{Jarzynski3} for work done under a time symmetric protocol. 
\section*{I. Complement to the Derivation of TUR}
Let us complement our proof of TUR outlined in the main text. 
First, we confirm that ${\rm C}_1$ and ${\rm C}$ are tangent to the line $-\Delta\beta J$. 
From the Cramer's theorem\cite{Ellis2}, the rate function is given as 
\begin{align}
&I(J)={\rm sup}_{\lambda}(\lambda J-\log\langle e^{\lambda J_1}\rangle). \tag{S1} \label{S1}
\end{align}
The extremal condition $\frac{d}{d\lambda}(\lambda J-\log\langle e^{\lambda J_1}\rangle)=0$ is equivalent to 
\begin{align}
&J-\frac{\langle Je^{\lambda J_1}\rangle}{\langle e^{\lambda J_1}\rangle}=0. \tag{S2}\label{S2}
\end{align}
Substituting $\lambda=-\Delta\beta$ and applying EFT, 
we can verify that supremum is achieved for $\lambda=-\Delta\beta$ from $\langle J_1e^{-\Delta\beta J_1}\rangle=-\langle J_1\rangle$ and $\langle e^{-\Delta\beta J_1}\rangle=1$, and  
we obtain ${\rm I}(-\langle J_1\rangle)=\Delta\beta\langle J_1\rangle$. 
For other values of $J$, the supremum is achieve for $\lambda(\neq-\Delta\beta)$, and therefore $-\Delta\beta J$ is actually tangent to $C_1$ and ${\rm C}$. 
Similarly, we can verify that the horizontal axis is tangent to ${\rm C}_2$ and ${\rm C}$. 

Then, the curve ${\rm C}$ is placed above the lines $-\Delta\beta J$ and $J=0$, and convexly connects the points ${\rm P}_1$ and ${\rm P}_2$. 
From the regularity condition, the rate function $I(J)$ should locally follows  
the evaluations (5) for $J\cong -\langle J_1\rangle$ and (6) for $J\cong\langle J_1\rangle$. This condition violates the convexity as illustrated in Fig. 1.
\setcounter{section}{0}
\setcounter{figure}{0}
\begin{figure}
\includegraphics[scale=0.5]{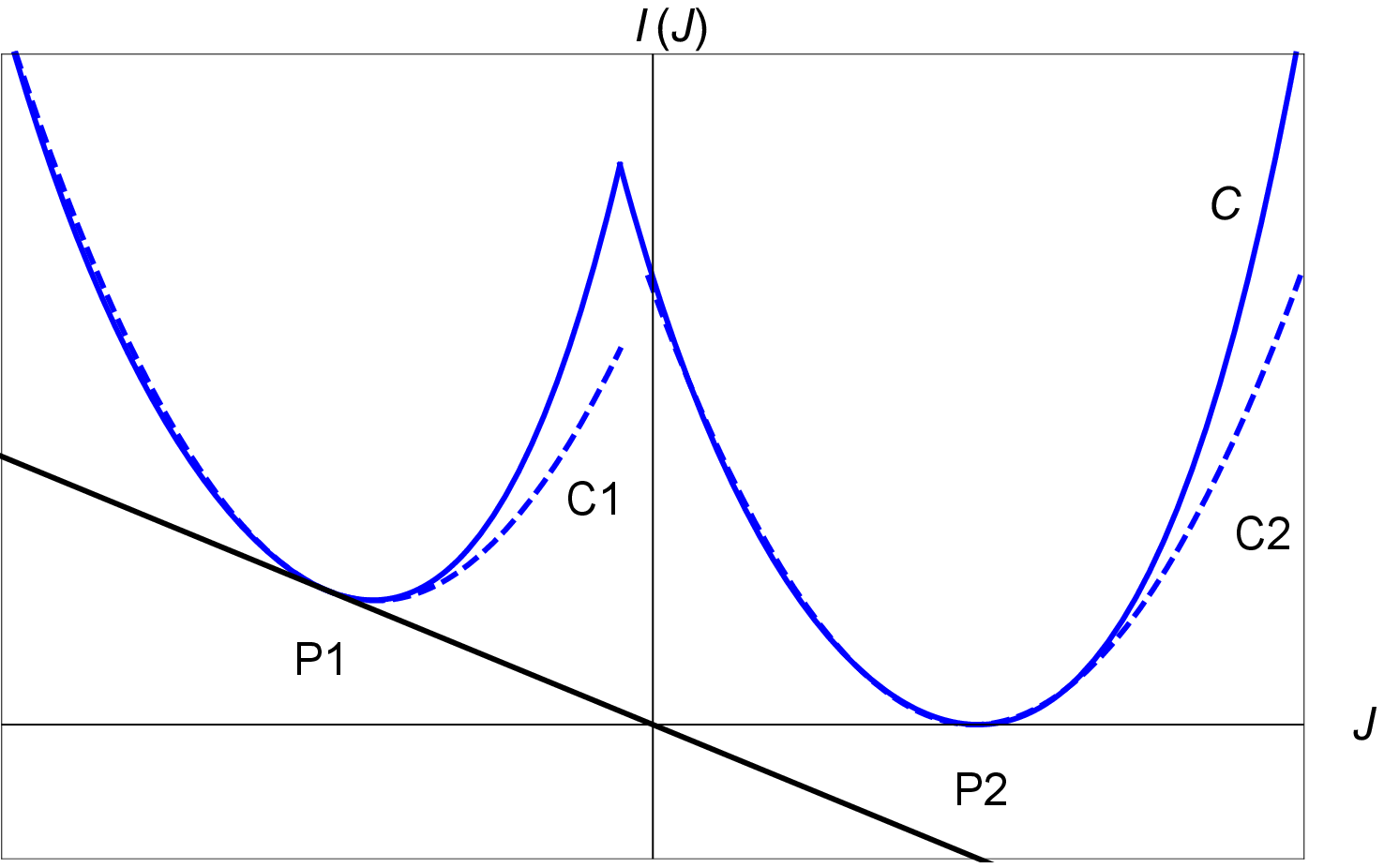}
\caption{Schematic illustration of a curve $C$ (blue) corresponding to the rate function $I(J)$ that locally follows ${\rm C}_1$ and ${\rm C}_2$. The curve ${\rm C}$ is tangent to the line $-\Delta\beta J$ (black) and $J=0$, and locally follows the evaluations from the central limit theorem ${\rm C}_1$ and ${\rm C}_2$ (dashed blue curves).}
\end{figure} 

Strictly speaking, the regularity condition requires that the curve ${\rm C}$ locally follows ${\rm C}_2$ near $J=\langle J_1\rangle$ so that ${\rm C}$ is placed above the common tangent line ${\rm L}$ of ${\rm C}_1$ and ${\rm C}_2$ in between the tangent points ${\rm Q}_1$ and ${\rm Q}_2$ of ${\rm L}$. 
In other words, we consider a class of systems that satisfy this requirement.

Let us confirm this requirement is actually satisfied for example if the convergence in (3) is sufficiently rapid, and (3) approximately holds for a relatively small number $N=N_c$. 
This also implies the rapid convergence for the central limit theorem.   
The requirement trivially holds for the case of Gaussian $p_\tau(J_1)$, since the curve ${\rm C}_1$ and ${\rm C}_2$ coincide, and ${\rm L}$ is tangent to ${\rm C}_3$. And, TUR holds irrespective of the value of $\Delta\beta$.      
For nonGaussian $p_\tau(J_1)$, the central limit theorem holds for $N=N_c$ copies with a sufficient accuracy $\epsilon(>0)$, i.e., the left hand side of (12) belongs to $[1-\epsilon,1+\epsilon]$ for (13). Here, the probability distribution of $\frac{J_{tot}}{N_c}$ for $N_c$ copies $p_\tau(\frac{J_{tot}}{N_c}=J)$ has a peak at $\frac{J_{tot}}{N_c}=\langle J_1\rangle$ with a variance $\frac{{\rm Var}[J_1]}{N_c}$.     
Then, there is a real number $\kappa$ of the same order as $\sqrt{\frac{N_c}{\langle J_1^2\rangle/\langle J_1\rangle^2-1}}$, which is considered as small from the smallness of $N_c$ and the assumption of the large fluctuation of the charge transfer in the derivation after (6),  and satisfies 
\begin{align}
&\kappa\sqrt{\frac{{\rm Var}[J_1]}{N_c}}\geq\langle J_1\rangle. \tag{S3}\label{S3}
\end{align}
Since we assumed that $\langle J_1\rangle\geq\frac{1}{2}\Delta\beta{\rm Var}[J_1]$, (\ref{S3}) implies that (12) holds for $J\leq J_c=\langle J_1\rangle-\frac{1}{2}\Delta\beta{\rm Var}[J_1]$ within the accuracy $\frac{1}{N_c}\log(1\pm\epsilon)$ by choosing $\kappa_1={\rm min}\{\kappa,\frac{1}{2}\sqrt{N_c\Delta\beta^2{\rm Var}[J_1]}\}$ in (13), where $J=J_c$ is the tangent point ${\rm Q}_2$ of ${\rm L}$ on ${\rm C}_2$. 
Therefore, the curve ${\rm C}$ actually follows ${\rm C}_2$ for $J\leq J_c$.     

Let us also explore the exceptional case of (8) where the regularity condition is slightly violated.    
In Fig. 2, we illustrate the ratio $\left(\frac{\langle J_1\rangle^2}{{\rm Var}[J_1]}\right)/\left(\frac{\sigma}{2}\right)$ as a function of $\Delta\beta$ for the family of irregular distributions generated by the minimal distribution (8) for $r\geq\frac{1}{10}$, whose supports are finite and has relatively small variances. 
\begin{figure}
\includegraphics[scale=0.5]{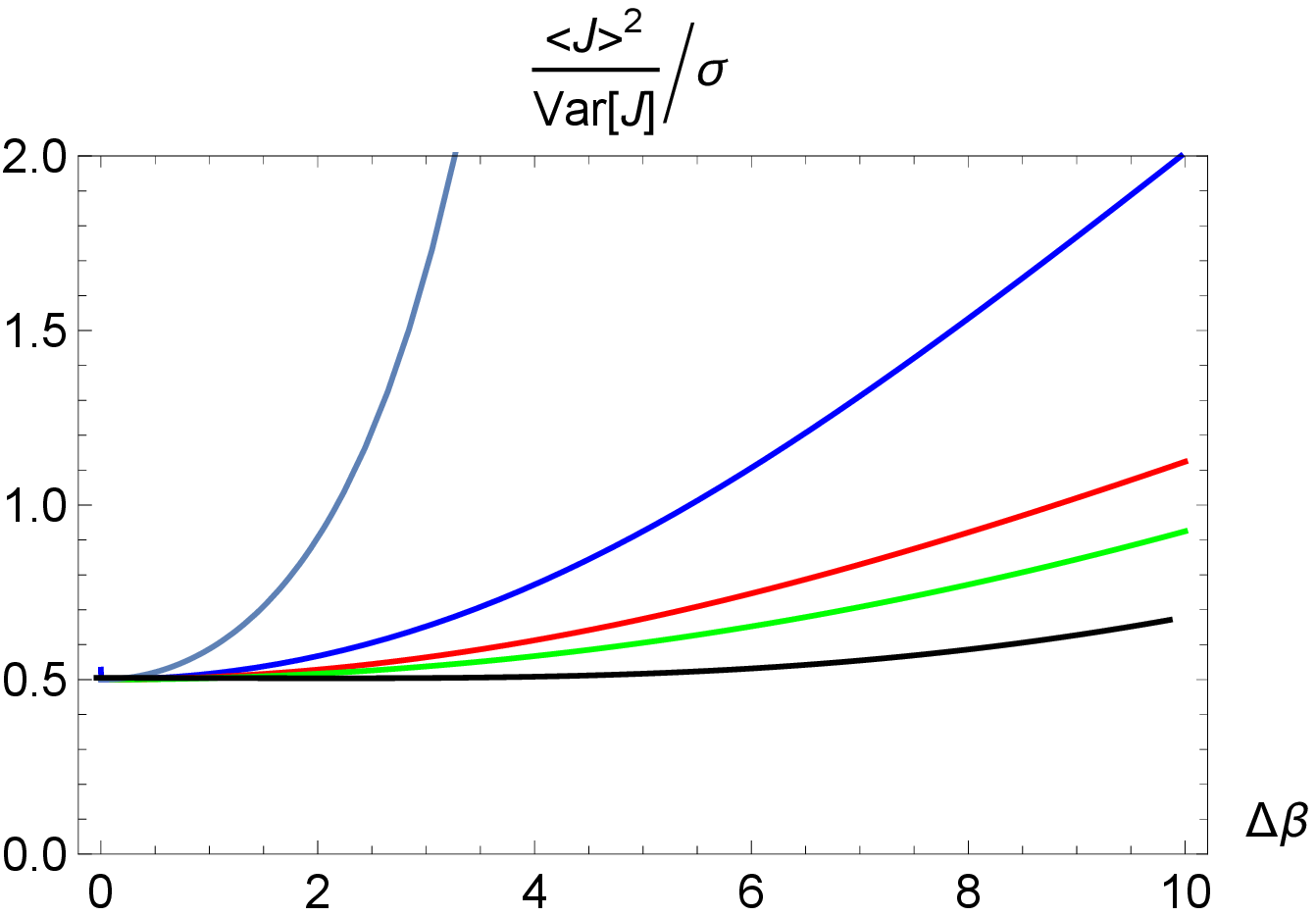}
\caption{The ratio $\frac{\langle J\rangle^2}{{\rm Var}[J]}/\sigma$ for $(1-(\frac{|J|}{a})^r)e^{\frac{\Delta\beta J}{2}}$
 for $r=0,1,2,\frac{1}{2},\frac{1}{10}$ ( blue, red, green, and black, respectively) and for the minimal distribution (light blue). Without loss of generality, we set $a=1$. The ratio is almost $2$ for small $\Delta\beta$ for all the cases, and smaller than $2$ for intermediate values of the affinity $\Delta\beta$ for $r=\frac{1}{10}$.}
\end{figure}   
Intuitively, the smallness of the variance amounts to the large curvature of the rate function at $J=\langle J\rangle$, and it is difficult for the curve $C$ to locally follow $C_2$. 
 
\section*{II. Derivation of TUR equality in short time limit}
Eq. (14) in the main text holds under the condition (13). 
Then, TUR equality (9) is derived by comparing (14) with EFT applied to the ensemble of the copies (15)
\begin{align}
\frac{p_\tau(J_{tot})}{p_\tau(-J_{tot})}&=e^{\frac{2\langle J_{tot}\rangle}{{\rm Var}[J_{tot}]}J_{tot}} \nonumber \\
&=e^{\Delta\beta J_{tot}}. \tag{S4}\label{S4}
\end{align}
Actually, we obtain from (\ref{S4}) 
\begin{align}
\frac{\langle J_1\rangle}{{\rm Var}[J_1]}=\frac{\Delta\beta}{2}, \tag{S5} \label{S5}
\end{align}  
where we used $\langle J_{tot}\rangle=N\langle J_1\rangle$ and ${\rm Var}[J_{tot}]=N{\rm Var}[J_1]$ in (\ref{S4}).
Therefore, we explore the range of validity of (14). 
In the short time limit, the amount of the charge transfer $\langle J_1\rangle$ during $\tau$ is a  decreasing function of $\tau$. 
Similarly, $\langle J_1\rangle$ is supposed to be an increasing function of $\Delta\beta$. 
Thus, we assume that the mean and variance of the charge transfer satisfy the asymptotic expressions (16) and (17). 
Eq. (14) holds provided that the condition (13) is satisfied. 
We also specify the entropy production
\begin{align}
\Delta\beta\langle J_1\rangle=\frac{\kappa_3}{N^\delta} \tag{S6}\label{S6}
\end{align}   
with an unknown exponent $\delta$ and an ${\cal O}(1)$ dimensionless constant $\kappa_3$. 
Substituting (16) and (17) into $|N\langle J_1\rangle|\leq\kappa_1\sqrt{N{\rm Var}[J_1]}$, and (17) into (\ref{S4}), we obtain  
\begin{align}
(\frac{\tau}{\tau_0})^{p-\frac{1}{2}q}(\frac{\Delta\beta}{\Delta\beta_0})^{\alpha-\frac{1}{2}\gamma}&=\frac{\kappa_1\sqrt{K_2}}{\sqrt{N}K_1} \tag{S7}\label{S7} \\
(\frac{\tau}{\tau_0})^p(\frac{\Delta\beta}{\Delta\beta_0})^{\alpha+1}&=\frac{\kappa_3}{\Delta\beta_0 K_1 N^\delta}. \tag{S8}\label{S8}
\end{align}
Eqs. (\ref{S7}) and (\ref{S8}) are rewritten as
\begin{align}
&\frac{q}{2}-p\frac{\frac{\gamma}{2}+1}{\alpha+1}=0 \nonumber \\
&\frac{1}{2}+\delta(\frac{\frac{\gamma}{2}+1}{\alpha+1}-1)=0 \nonumber \\
&(\frac{\kappa_3}{\Delta\beta_0 K_1})^{\frac{\frac{\gamma}{2}+1}{\alpha+1}}=\frac{\kappa_3}{\kappa_1\Delta\beta_0\sqrt{K_2}}. \tag{S8}\label{S9}
\end{align}
On the other hand, the ratio of the mean $\langle J_1\rangle$ and variance ${\rm Var}[J_1]$ 
\begin{align}
\frac{\langle J_1\rangle}{{\rm Var}[J_1]}=\frac{K_1}{K_2}(\frac{\tau}{\tau_0})^{p-q}(\frac{\Delta\beta}{\Delta\beta_0})^{\alpha-\gamma} \tag{S10}\label{S10}
\end{align}
should be equal to $\frac{\Delta\beta}{2}$. 
These conditions are actually uniquely fulfilled by a solution
\begin{align}
p=q, \alpha=\gamma+1, \delta=1, \frac{K_1}{K_2}=\frac{\Delta\beta_0}{2}, \kappa_3=2\kappa_1^2. \tag{S11}\label{S11}
\end{align}
As a byproduct, (\ref{S10}) shows that the mean and variance of the charge transfer have the same scaling exponent $p$. 
This completes the derivation of (9), (10) and (11).

\section*{III. Concrete examples}
In this section, let us first consider the model described by the Hamiltonian (22).
The work done is calculated as 
\begin{align}
W=\int_0^{\frac{\tau}{2}}\frac{m\omega_0^2}{2\tau}q(t)^2 dt-\int_{\frac{\tau}{2}}^\tau\frac{m\omega_0^2}{2\tau}q(t)^2 dt \tag{S12}\label{S12}
\end{align}
under the initial distribution $\rho(q(0),p(0),0)=\frac{1}{Z(0)}e^{-\beta H(q(0),p(0),0)}$. 
Then, in Fig. 2, EFT (4) can be numerically confirmed for the probability distribution $p_\tau(W)$, which is calculated as the Fourier transformation of the characteristic function.
\begin{figure}
\includegraphics[scale=0.5]{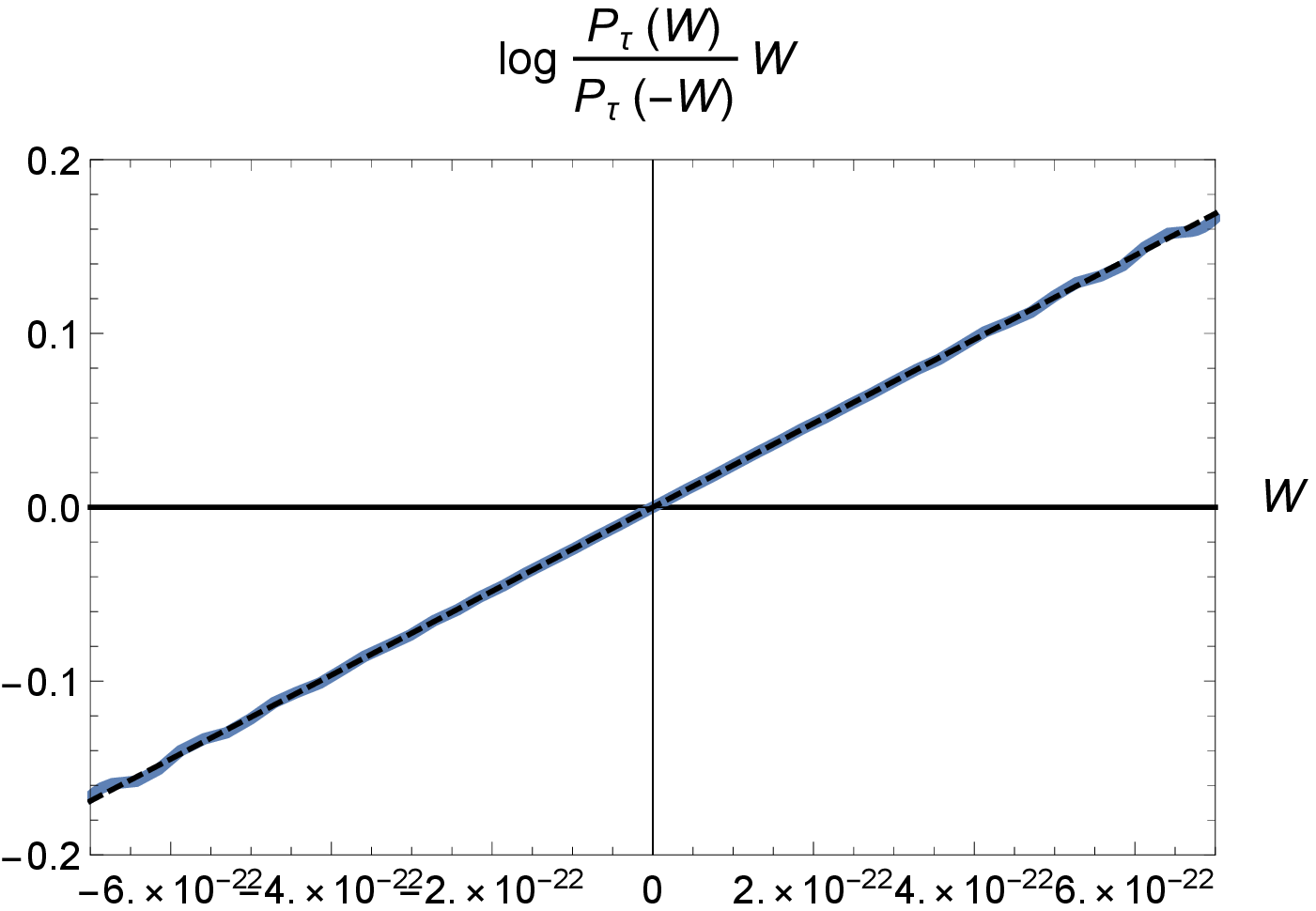}
\caption{Plot of $\log\frac{p_\tau(W)}{p_\tau(-W)}$(blue) and $\Delta\beta W$(dashed black line), which are in good agreement. The duration is $\tau=0.02{\rm s}$, the temperature is $T=300{\rm K}$, the mass $m=10^{-9}{\rm kg}$, and the spring constant is $m\omega_0^2=10^{-8}{\rm kg/s^2}$.}      
\end{figure}
We also investigate the power-law scaling of $\langle W\rangle\propto\tau^p$ with $p=2$ in Fig. 3. 
\begin{figure}
\includegraphics[scale=0.5]{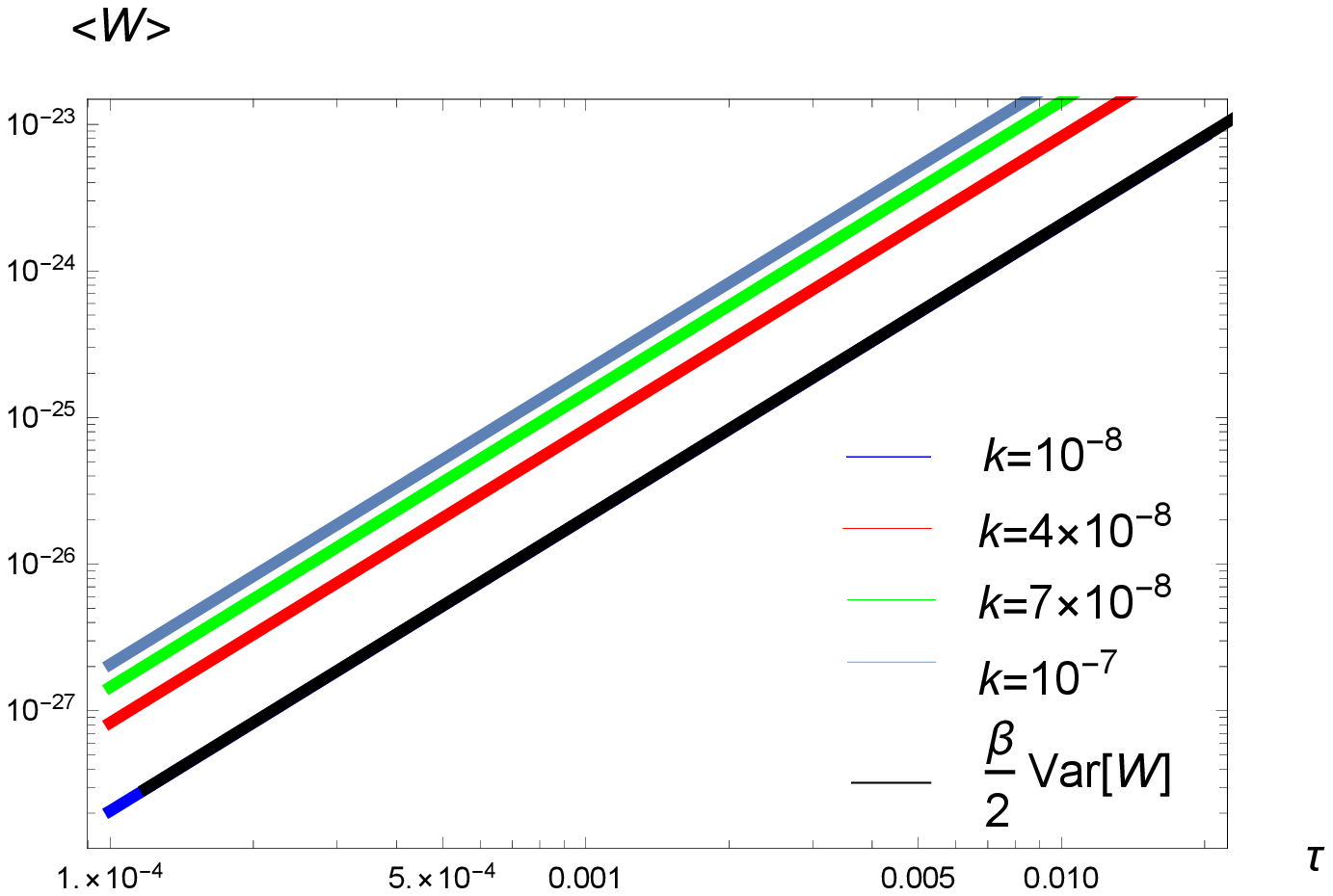}
\caption{Log-log plot of $\langle W\rangle$ as a function of $\tau$. The parameters are the same except for the spring constant $k=m\omega_0^2$. The relation $\langle W\rangle\propto\tau^2$ is numerically confirmed for all the values of spring constant $k=10^{-8}, 4\times 10^{-8}, 7\times 10^{-8}$ and $10^{-7}{\rm kg/s^2}$.}
\end{figure}      

In addition to the verification of TUR in short time regime illustrated in Fig. 1 in the main text, we plot the ratio $\frac{\langle W\rangle^2}{{\rm Var}[W]}/\frac{\sigma}{2}$ for relatively long time regime in Fig. 4. 
The ratio $\frac{\langle W\rangle^2}{{\rm Var}[W]}/\frac{\sigma}{2}$ shows a non-monotonic complicated dependence on $\tau$. 
We can verify that TUR holds in the entire region. %Also, (\ref{TPE4}) holds with a relatively small $\varepsilon$ in particular for $\tau\geq 1{\rm s}$.    
\begin{figure}
\includegraphics[scale=0.5]{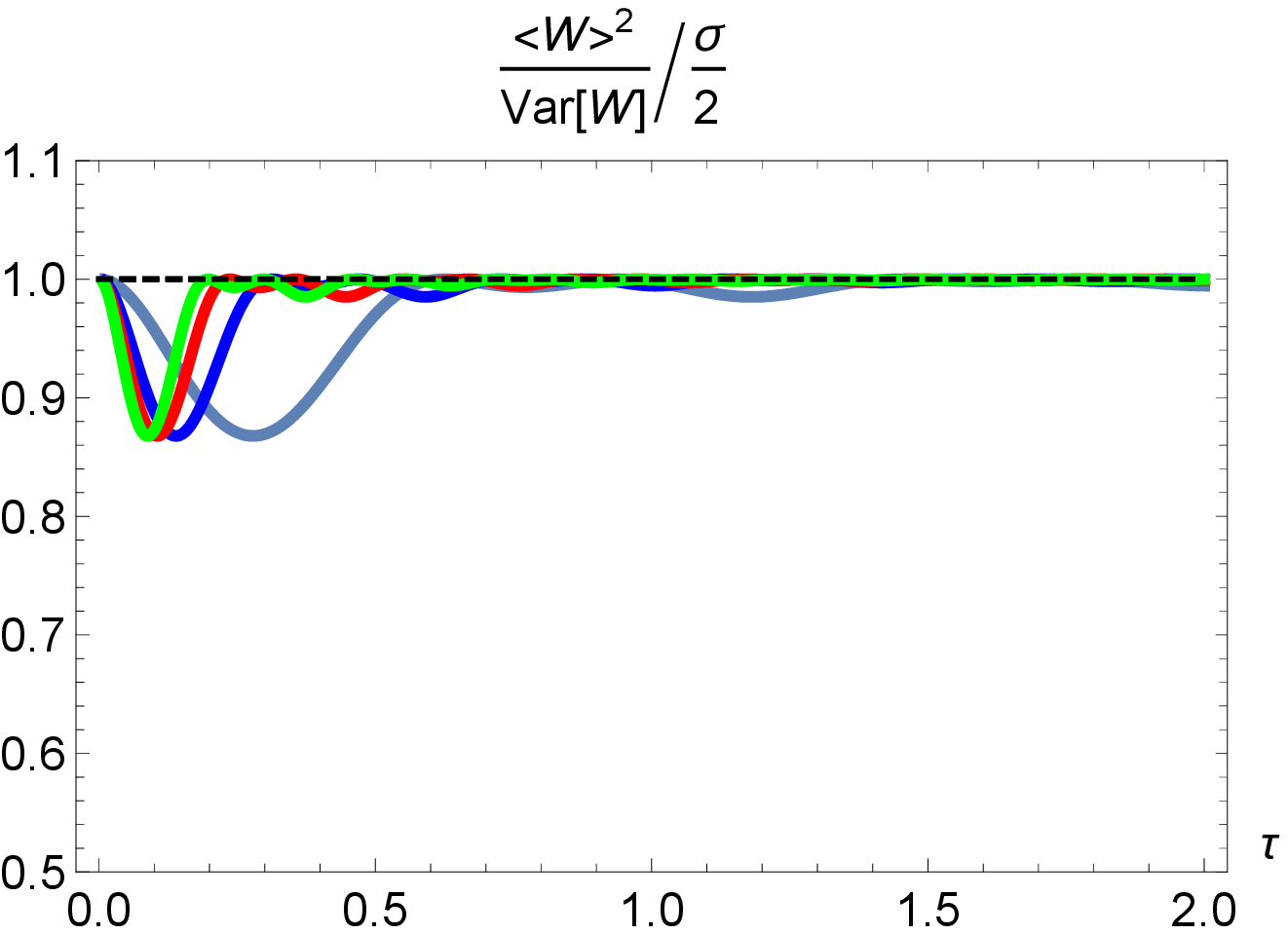}
\caption{The plot of $\frac{\langle W\rangle^2}{{\rm 
Var}[W]}/\frac{\sigma}{2}$ as a function of $\tau$. The parameters are   
$T=300{\rm K}$ and $m=10^{-9}{\rm Kg}$, and $m\omega_0^2$ is changed from 
$10^{-8}{\rm kg/s^2}$ to $10^{-7}{\rm kg/s^2}$ by a step $3\times 10^{-8}{\rm 
kg/s^2}$ (light blue, blue, red, and green curves). The dashed line indicates 
the equality condition of TUR.}
\end{figure}

{\it Example 2.---} 
As an experimental design, we consider another example described by an overdamped Langevin equation.  
In this manner, we explain that the work distribution of an externally dragged Brownian particle\cite{Wang2,Zon1,Tasaki1,Monnai2} also satisfies TUR for work done. 
Here, the particle is immersed in an environment at an inverse temperature $\beta$. Since the temperature gradient is absent, the driving force is the external dragging for example by optical tweezers. Therefore, the work done is relevant for this system.       
For simplicity, we consider the one-dimensional case described by a Markovian  overdamped Langevin equation with a moving harmonic potential
\begin{align}
\gamma\dot{x}(t)=-k(x(t)-f(t))+\xi(t), \tag{S13}\label{Langevin1}
\end{align}
where $x(t)$, $\gamma$, $k$, and $f(t)$ denote the position of the particle, the friction coefficient, the spring constant, and the position of the center of the potential, respectively. 
The thermal noise $\xi(t)$ stands for the Gaussian stochastic process satisfying the fluctuation-dissipation relation of the second kind
\begin{align}
\langle f(t)f(t')\rangle=2\frac{\gamma}{\beta}\delta(t-t'). \tag{S14}\label{correlation1}  
\end{align}
%where $\beta$ denotes the inverse temperature. 
The system is initially in thermal equilibrium, and without loss of generality the center of the potential is placed in the origin at $t=0$.  
Then, the center of the potential starts to move. 
The work done on the system during $0\leq t\leq\tau$ is given by\cite{Wang2,Zon1,Monnai2}
\begin{align}
W=-k\int_0^\tau(x(t)-f(t))\dot{f}(t)dt, \tag{S15}\label{work2}
\end{align}
which is the change of the internal energy due to the change of the control parameter $f(t)$. 
Since Eq. (\ref{Langevin1}) is linear and $\xi(t)$ is Gaussian, the distribution function of $W$ obeys a normal distribution. Therefore, we can skip preparing the copies to use the central limit theorem and the large deviation principle.   
By exactly solving Eq. (\ref{Langevin1}), we can verify that the probability distribution of work done $p_\tau(W)$ satisfies FT\cite{Zon1}
\begin{align}
\frac{p_\tau(W)}{p_\tau(-W)}=e^{\beta W}. \tag{S16}\label{fluctuation5}
\end{align} 
To explore TUR, note that the mean and the variance of the work done are
\begin{align}
&\langle W\rangle=k\int_0^\tau dt_1\int_0^{t_1}dt_2e^{-\frac{k}{\gamma}(t_1-t_2)}\dot{f}(t_1)\dot{f}(t_2) \nonumber \\
&{\rm Var}[W]=\frac{2}{\beta}\langle W\rangle. \tag{S17}\label{cummulant2}
\end{align}
Hence, TUR equality (9) holds for the work done for an arbitrary $\tau$ 
\begin{align}
\frac{\langle W\rangle^2}{{\rm Var}[W]}=\frac{\sigma}{2}, \tag{S18}\label{TPE5}
\end{align}     
where $\sigma=\beta \langle W\rangle$ is called entropy production\cite{Wang2}. 
For this example, the equality condition of TUR holds for arbitrary duration $\tau$ and driving protocol $f(t)$.  
As for the scaling properties (10) and (11), the exponents are common, especially $\alpha=0$ in general.  The exponent of $\tau$ is equal to $p=2$ for the case of the constant velocity $\dot{f}(t)=v$.    

From thermodynamic point of view, the dissipative work divided by the temperature $\beta(W-\Delta F)$ is positive-semidefinite when averaged and is usually identified as the entropy production for isothermal processes. 
Here, the driving protocol $f(t)$ is arbitrary and not necessarily symmetric. 
We can rationalize this point by noting that the free energy change $\Delta F$ is negligible for the present case with a fixed $k$. For this reason, we can identify $\beta W$ as the dissipative work divided by the temperature. Also, the mean $\langle\beta W\rangle$ is nonnegative.  
\section*{IV. Exchange fluctuation theorem for work done}
In this section, we derive EFT for work done under a time symmetric protocol.
For this purpose, let ${\bf z}$ denote the point of the phase space, i.e., a set of positions and momenta of the system. 
Here, $\lambda(t)=\lambda(\tau-t)$ ($0\leq t\leq\tau$) stands for the time symmetric control parameter so that the probability distributions for time forward and reversed processes are identical under the same initial condition. 
The work done $\hat{W}({\bf z}(0))=\int_0^\tau\frac{\partial H({\bf z}(t),\lambda(t))}{\partial\lambda(t)}\dot{\lambda}(t)$ satisfies FT (see \cite{Crooks3} for stochastic systems). 

We define the time reversed trajectory $\overline{{\bf z}}(t)={\bf z}^*(\tau-t)$, and assume the 
time reversal invariance of the Hamiltonian
\begin{align}
H({\bf z}^*(t),\lambda_k(t))=H(\overline{{\bf z}}(\tau-t),\lambda(t)), \tag{S19}\label{invariance1}
\end{align}
where the symbol $*$ stands for the time reversal. 
The initial state ${\bf z}(0)$ is sampled from the canonical ensemble
\begin{align}
P({\bf z}(0))=\frac{1}{Z}e^{-\beta H({\bf z}(0)),\lambda(0))}, \tag{S20}\label{initial1}
\end{align}
where $Z$ denotes the partition function. 
Then, for a trajectory ${\bf z}(t)$, the ratio of the probability distribution (\ref{initial1}) for the time forward and reversed processes is given by 
\begin{align}
\frac{P({\bf z}(0))}{P(\overline{{\bf z}}(0))}=e^{\beta\hat{W}({\bf z}(0))}, \tag{S21}\label{ratio1}
\end{align}
where $\hat{W}({\bf z}(0))$ is the work done on the system. 
The work done is odd under the time reversal
 $\hat{W}({\bf z}(0))=-\hat{W}_{tot}^*(\overline{{\bf z}}(0))$. 
Combined with (\ref{ratio1}), the probability distribution of the work done on the total system satisfies
\begin{align}
&p_\tau(W) \nonumber \\
&=\int d{\bf z}(0)P({\bf z}(0))\delta(W-\hat{W}({\bf z}_{tot}(0))) \nonumber \\
&=\int d\overline{{\bf z}}(0)e^{\beta\hat{W}(0)}P(\overline{{\bf z}}(0))\delta(W+\hat{W}(\overline{{\bf z}}(0)) \nonumber \\
&=e^{\beta W}p_\tau(-W). \tag{S22}\label{fluctuation1}
\end{align}  

\end{widetext}